\documentstyle[sprocl]{article}

\advance\textheight by \topskip
\begin{document}
\def\lsim{\mathrel{\lower2.5pt\vbox{\lineskip=0pt\baselineskip=0pt
\hbox{$<$}\hbox{$\sim$}}}}
\def\gsim{\mathrel{\lower2.5pt\vbox{\lineskip=0pt\baselineskip=0pt
\hbox{$>$}\hbox{$\sim$}}}}

\def\gs{SU(2)_{\rm L} \times U(1)_{\rm Y}}
\def\mt{{\tilde{m}^2}_{{\scriptscriptstyle t}}}
\def\mb{{\tilde{m}^2}_{{\scriptscriptstyle b}}}
\def\mz{{m}_{\scriptscriptstyle Z}^{2}}
\def\at{{A_{\scriptscriptstyle t}}}
\def\ab{{A_{\scriptscriptstyle b}}}
\def\qt{{Q_{\scriptscriptstyle t}}}
\def\qb{{Q_{\scriptscriptstyle b}}}
\def\tu{{\tilde{t}_{{\scriptscriptstyle 1}}}}
\def\td{{\tilde{t}_{{\scriptscriptstyle 2}}}}
\def\bu{{\tilde{b}_{{\scriptscriptstyle 1}}}}
\def\bd{{\tilde{b}_{{\scriptscriptstyle 2}}}}
\def\tauu{{\tilde{\tau}_{{\scriptscriptstyle 1}}}}
\def\taud{{\tilde{\tau}_{{\scriptscriptstyle 2}}}}
\def\mnu{{\tilde{m}^2_{{\scriptscriptstyle \nu}}}}
\def\mtau{{\tilde{m}^2}_{{\scriptscriptstyle \tau}_{\scriptscriptstyle 1}}}
\def\mtad{{\tilde{m}^2}_{{\scriptscriptstyle \tau}_{\scriptscriptstyle 2}}}
\def\w{{\tilde{W}^{\pm}}}
\def\h{{\tilde{H}^{\pm}}}
\def\chiu{{\tilde{\chi}^{+}}_{{\scriptscriptstyle 1}}}
\def\chid{{\tilde{\chi}^{+}}_{{\scriptscriptstyle 2}}}
\def\sw{s_{{\scriptscriptstyle W}}}
\def\cw{c_{{\scriptscriptstyle W}}}
\def\mfa{{\tilde{m}^2}_{{\scriptscriptstyle fa}}}
\def\mfb{{\tilde{m}^2}_{{\scriptscriptstyle fb}}}
\def\mi{{\tilde{M}^2}_{{\scriptscriptstyle i}}}
\def\mj{{\tilde{M}^2}_{{\scriptscriptstyle j}}}
\newcommand{\sfe}{{\tilde f}}
\newcommand{\sg}{{\tilde \chi}}
\newcommand{\sne}{{\tilde \chi^o}}
\newcommand{\scp}{{\tilde \chi^+}}
\newcommand{\dx}{dx}  
\newcommand{\dy}{dy}  
\newcommand{\slas}[1]{\rlap/ #1}
\newcommand{\diag}{{\rm diag}}
\newcommand{\Tr}{{\rm Tr}}
\vspace*{-1cm}
\begin{flushright}
{November 1997}\\
{hep-ph/9711441}
\end{flushright}
\begin{center}
\begin{large}
\begin{bf}
DECOUPLING OF SUPERSYMMETRIC PARTICLES IN THE MSSM\\
\end{bf}
\end{large}
\vspace{0.4cm}
ANTONIO DOBADO\\
\vspace{0.1cm} 
{\em  Departamento de F{\'\i}sica Te{\'o}rica\\
  Universidad Complutense de Madrid\\
 28040-- Madrid,\ \ Spain} \\
\vspace{0.2cm} 
 MARIA J. HERRERO$^{\dagger}$\\
\vspace{0.1cm}
and\\
\vspace{0.1cm}
SIANNAH PE{\~N}ARANDA\\
\vspace{0.1cm}
{\em  Departamento de F{\'\i}sica Te{\'o}rica\\
  Universidad Aut{\'o}noma de Madrid\\
  Cantoblanco,\ \ 28049-- Madrid,\ \ Spain}
\end{center}
\begin{center}
{\bf ABSTRACT}
\end{center}
\begin{quotation}
\noindent
A heavy supersymmetric spectrum at the Minimal Supersymmetric Standard Model is
considered and the decoupling from the low energy electroweak scale is analyzed. A 
formal and partial proof of decoupling of supersymmetric particles in the limit where their 
masses are larger than the electroweak scale is performed  by integrating out all the 
sparticles to one loop and by evaluating the effective action for the 
standard electroweak gauge bosons $W^{\pm}, Z$ and $\gamma$. The Higgs sector is not considered
here. Analytical results for the two-point functions 
of the electroweak gauge bosons and the $S, T$ and $U$ parameters, to be valid in that limit, are 
also presented.  A discussion on how the decoupling takes place in terms of both the 
physical sparticle masses and the non-physical mass parameters as the $\mu$-parameter and the 
soft-breaking parameters is included.
\end{quotation}
---------------------------------------------------------------------------\\
${\dagger}${\em  Talk presented at the International Workshop on quantum effects in the MSSM,
 Barcelona, 9-13 September, 1997.}\\

\section{Introduction}
\label{sec:mssm}
\renewcommand\baselinestretch{1.3}
\hspace*{0.5cm}At present, there are indications that when the spectrum of supersymmetric (SUSY) particles 
at the Minimal Supersymmetric Standard Model (MSSM) is considered
much  heavier  than the low energy electroweak scale they decouple from the low energy physics, even at the quantum level, and 
the resulting low energy effective theory is the Standard Model (SM)  itself. However, a rigorous proof 
of decoupling is still lacking. On one hand there are numerical studies of 
observables that measure electroweak radiative corrections, like $\Delta r$ and $\Delta \rho$
 \cite{CHA}, or the $S, T$ and $U$ parameters \cite{HA2} as well as in the $Z$ boson, top quark and Higgs 
decays \cite{SOLA}, which indicate  that the one loop corrections from supersymmetric particles 
decrease up to negligible values in the limit of very heavy sparticle masses. Decoupling of SUSY
particles is also found in some analytical studies of these and 
 related observables~\cite{CHA,GL1}, as well as some computations of the effective potential
 for the scalar sector~\cite{QUI}, in the asymptotic limits where some of the SUSY mass 
parameters are considered infinitely heavy. 
 
 The question whether the Decoupling Theorem \cite{OJO} applies or not in the case of heavy 
sparticles in MSSM \cite{HAMSSM} is not obvious at all, in our opinion. The MSSM is a gauge theory which
incorporates the spontaneous symmetry breaking $\gs \rightarrow U(1)_{\rm em}$ and chiral 
fermions as the SM 
and  therefore, the direct application of the Decoupling Theorem should, in the
principle, be questioned~\cite{ANT}. Examples  where the 
Decoupling Theorem does not hold are well known. Particularly interesting 
are the cases of the Higgs particle and the top quark in the SM which are known
not to decouple from low energy physics \cite{VE,CHI,MJ}. 

In our opinion, a formal proof of decoupling must involve the explicit 
computation of the effective action by integrating out one by one all the 
sparticles in the MSSM to all orders in perturbation theory, and by considering 
the heavy sparticle masses limit. The proof will be conclusive if the remaining   
effective action, to be valid at energies much lower than the supersymmetric particle 
masses, turns out to be that of the SM with all the SUSY effects being absorbed into  
a redefinition of the SM parameters or else they are suppressed by inverse powers  
of the SUSY particle masses and vanish in the infinite masses limit.

In the present work we discuss part 
of the effective action which results by integrating out all the SUSY particles of the MSSM,  
except the Higgs sector, at the one loop level. This is a reduced version of a more complete 
paper to which we refer the reader for a more detailed discussion~\cite{GEISHA} . The integration of 
the Higgs sector will be consider separately in a forthcoming 
work \cite{DMS}. The part of the effective action we have chosen to start with is the one for the 
electroweak gauge bosons, $W^{\pm}$, Z and $\gamma$ and, in 
particular, we have devoted more attention on the derivation of the two point functions with external
 $W^{\pm}$, Z and $\gamma$ gauge bosons. This will allow us to derive, in addition, the contributions  
from the SUSY particles to the $S, T$ and $U$ parameters in the large SUSY masses limit and
 to conclude how the decoupling
really occurs in these parameters. In order to keep our computation of the heavy 
SUSY particle quantum effects in a general form we have chosen to work with the masses 
themselves. They are the proper parameters of the large mass expansions instead of another more model 
dependent choices as the $\mu$-parameter or the soft-SUSY-breaking parameters, 
$M_{{\scriptscriptstyle {\tilde Q}}}, M_{{\scriptscriptstyle {\tilde U}}}, 
M_{{\scriptscriptstyle {\tilde D}}}, M_{{\scriptscriptstyle {\tilde L}}},
M_{{\scriptscriptstyle {\tilde E}}}$, and $M_{1}, M_{2}$ 
\footnote{We use here the same notation as in ~\cite{HAMSSM,GEISHA,HC}}
.We have considered the physicaly plausible situation where all the sparticle masses are large as compared to 
the electroweak scale but they are allowed, in principle, to be different from each other. 
We will explore the interesting question of to what extent the usual hypothesis of SUSY masses being  
generated by soft-SUSY- breaking terms and the universality of the mass parameters
do or do not play a relevant role  in getting decoupling. In fact, we will show in this
paper, that the basic requirement of $\gs$ gauge invariance on the SUSY breaking 
terms is sufficient to obtain decoupling in the MSSM.

Finally, we have dedicated special attention and have been very careful in evaluating 
analytically the large SUSY masses limit of the Green functions. For this purpose, 
we have applied the so-called m-Theorem \cite{GMR} which provides a rigorous technique to compute 
Feynman integrals with both large and small masses in the asymptotic regime of the large 
masses being very heavy. 

 The paper is organized as follows: In section 2 we present a brief discussion on how to get 
large mass values for all the squarks, sleptons, neutralinos and charginos at the MSSM. The third 
section is devoted to present the effective action for the 
electroweak gauge bosons  $W^{\pm}$, Z and $\gamma$ in the MSSM that results by integrating 
out, in the path integral, squarks, sleptons, charginos and neutralinos to one-loop.
The asymptotic results in the large SUSY masses limit for the 
$\Sigma^{{\scriptscriptstyle XY}}$ and $R^{{\scriptscriptstyle XY}}$
functions are also included and analyzed. The decoupling of heavy sparticles in the $S, T$ 
and $U$ parameters is analyzed in the section 4. Explicit formulae for these parameters in the large 
SUSY masses limit as well as a discussion on these results are also presented 
in this section. Finally, the conclusions are summarized in section 5.

\section{Heavy supersymmetric spectrum at the MSSM}
\renewcommand\baselinestretch{1.3}

\hspace*{0.5cm} In this section we consider the mass eigenstates of the MSSM. Any set of particles of 
a given spin, baryon number, lepton number and the same $SU(3)_{\rm c} \times U(1)_{\rm em}$ quantum numbers can mix. 
Therefore, in principle, there can be mixing in all the sectors of the MSSM and one
must diagonalize mass matrices to obtain the mass eigenstates and the corresponding 
eigenvalues \cite{HAMSSM,HC}. We consider here all the sectors, except the Higgs sector that we prefer 
to analyze elsewhere \cite{DMS}.

In the present work we are interested in the Green functions with external electroweak gauge
bosons and in the large mass limit of the SUSY particles, which means the situation
where all the sparticle masses are much larger than the electroweak scale and the external
momenta. In particular this could be the case if the sparticle masses are well above 
$m_{\scriptscriptstyle Z}, m_{\scriptscriptstyle W}$ and $m_{\scriptscriptstyle t}$ 
but still below the few TeV upper bound that is imposed by the standard solution of the 
hierarchy problem. Furthemore, unless we are in a particular model, the masses of the 
various sparticles are, in general, different and independent. Therefore, we must take these 
masses to be large as compared to the external gauge boson masses and external momenta, but we must 
specify, in addition, how they compare to each other. More specifically, we assume here the most 
plausible situation where all the 
sparticle masses are large but close to each other; namely 
$\tilde{m}_{\scriptscriptstyle i}^{2}, \tilde{m}_{\scriptscriptstyle j}^{2} \gg 
M _{\scriptscriptstyle EW}^{2}, k^{2}$ and
$|\tilde{m}_{\scriptscriptstyle i}^{2}-\tilde{m}_{\scriptscriptstyle j}^{2}| \ll
|\tilde{m}_{\scriptscriptstyle i}^{2}+\tilde{m}_{\scriptscriptstyle j}^{2}|$, where 
$M _{\scriptscriptstyle EW}$ denotes any of the electroweak masses involved 
$(m_{\scriptscriptstyle Z}, m_{\scriptscriptstyle W}, m_{\scriptscriptstyle t},
\ldots)$ and $k$ denotes any external momentum. Notice that this includes the case that has
been the most studied in the literature where universality of sparticle masses is assumed.

This masses hypothesis, together with the requirement that all the sparticles must be heavier 
than their corresponding partners, imply some constraints on the SUSY parameters~\cite{GEISHA}. In
particular, in the squarks sector, if we ignore mixing between different generations to avoid 
unacceptable large flavor changing neutral currents and if we use the notation of the third family
for the mass eigenstates $\tu,\td,\bu,\bd$ and the corresponding mass squared
eigenvalues by $\tilde{m}^2_{{\scriptscriptstyle t}_{1,2}}$,
$\tilde{m}^2_{{\scriptscriptstyle b}_{1,2}}$, the previous conditions imply the following
constraints on the soft SUSY breaking and $\mu$ parameters:
\begin{equation}
\label{eq:rest1}
M_{{\scriptscriptstyle {\tilde Q}}}^{2}, M_{{\scriptscriptstyle {\tilde U}}}^{2} 
\gg m_{t}^{2}, m_{{\scriptscriptstyle Z}}^{2}\,\,, \hspace*{0.3cm}
|M_{{\scriptscriptstyle {\tilde Q}}}^{2}-M_{{\scriptscriptstyle {\tilde U}}}^{2}| \ll
|M_{{\scriptscriptstyle {\tilde Q}}}^{2}+M_{{\scriptscriptstyle {\tilde U}}}^{2}|\,\,,
\hspace*{0.3cm} m_{\scriptscriptstyle t}^{2} (\at-\mu \cot{\beta})^{2} < 
M_{{\scriptscriptstyle {\tilde Q}}}^{2} M_{{\scriptscriptstyle {\tilde U}}}^{2}\,.
\end{equation}
Here $\at$ is the trilinear coupling and $\cot{\beta} \equiv v_{1}/v_{2}$. The first 
condition implies, in turn, the limiting behaviour 
$\tilde{m}_{t_{1}}^{2}\rightarrow M_{{\scriptscriptstyle {\tilde Q}}}^{2}$\hspace*{0.1cm},
$\tilde{m}_{t_{2}}^{2}\rightarrow M_{{\scriptscriptstyle {\tilde U}}}^{2}$ \hspace*{0.1cm}. The
second condition means that $M_{{\scriptscriptstyle {\tilde Q}}}$ and 
$M_{{\scriptscriptstyle {\tilde U}}}$ must be close to each other and the third one means 
that the mixing can never be large. Similar conclusions can be obtained for the sbottoms.

In summary, in other to get large stop and sbottom masses one needs large 
values of the SUSY breaking masses $M_{{\scriptscriptstyle {\tilde Q}}},
M_{{\scriptscriptstyle {\tilde U}}}$ and $M_{{\scriptscriptstyle {\tilde D}}}$ as 
compared to the electroweak scale and, in order not to get a too large mixing, the  trilinear couplings
 $\at, \ab$ and the $\mu$ parameter must be constrained from above by the previous inequalities.
Notice that an arbitrarily large $\mu$ or $\at, \ab$ with $M_{{\scriptscriptstyle {\tilde Q}}},
M_{{\scriptscriptstyle {\tilde U}}},M_{{\scriptscriptstyle {\tilde D}}}$ fixed is not
allowed. 

Similar analysis can be done in the sleptons sector. In this sector $\tilde{\nu}, \tauu,
\taud$ are the mass eigenstates and the mass squared eigenvalues are $\mnu$, $\mtau$, $\mtad$ 
respectively. In this case, we conclude that
large squared sparticles masses, such that their sum be larger than their
difference, implies that $M_{{\scriptscriptstyle {\tilde L}}}^{2}$ and
$M_{{\scriptscriptstyle {\tilde E}}}^{2}$ are large, satisfying also 
$|M_{{\scriptscriptstyle {\tilde L}}}^{2}-M_{{\scriptscriptstyle {\tilde E}}}^{2}| \ll
|M_{{\scriptscriptstyle {\tilde L}}}^{2}+M_{{\scriptscriptstyle {\tilde E}}}^{2}|$. Neither
 $\mu$ nor $A_{\scriptscriptstyle \tau}$ can be taken arbitrarily large with 
$M_{{\scriptscriptstyle {\tilde L}}}, M_{{\scriptscriptstyle {\tilde E}}}$ fixed.

Concerning to the inos sector and by following the standard notation
~\cite{HAMSSM,GEISHA,HC}, we have denoted by $\chiu$, $\chid$ the 4-component Dirac fermions that 
represent the two physical charginos and by $\tilde{M}_{\scriptscriptstyle 1,2}^{+}$ their 
corresponding mass eigenvalues. The 4-component Majorana fermions which 
represent the 4 neutralinos are denoted by ${\tilde {\chi}}_{1}^{\scriptscriptstyle 0}, {\tilde {\chi}}_{2}^{\scriptscriptstyle 0},
{\tilde {\chi}}_{3}^{\scriptscriptstyle 0}$ and 
${\tilde {\chi}}_{4}^{\scriptscriptstyle 0}$ and their corresponding mass eigenvalues are
$\tilde{M}_{\scriptscriptstyle 1,2,3,4}^{0}$. 

In principle, the eigenvalues in the inos sector can be either positive or negative. We choose 
the SUSY breaking parameters $M_{1}$ and $M_{2}$ to be positive and allow
$\mu$ to be either positive or negative. The physical masses, 
$|\tilde{M}_{\scriptscriptstyle 1,2}^{+}|$ and 
$|\tilde{M}_{\scriptscriptstyle 1,2,3,4}^{0}|$ are, of course, positive \cite{GEISHA}. In case 
of negative eigenvalues we proceed following the method described in 
the second paper of~\cite{HC}. 

Notice that to reach the large SUSY masses limit that we are interested in, it is necessary to 
consider the mass parameters in the chargino sector in the range 
$M_{\scriptscriptstyle 2},\mu \gg m_{\scriptscriptstyle W}$
and therefore, to a very good approximation, the mixing is small and $\chiu $ will
be predominantly gaugino with a mass close to $M_{\scriptscriptstyle 2}$, whereas $\chid$ 
will be predominantly Higgsino with a mass close to $|\mu|$.

Analogously, in the large SUSY masses limit important simplifications do occur in the neutralino
sector~\cite{GEISHA}. In order to get the four 
neutralino masses larger than the electroweak scale it is necessary to consider the mass 
parameters in the range $M_{\scriptscriptstyle 1},M_{\scriptscriptstyle 2},\mu \gg
m_{\scriptscriptstyle Z}$. Therefore, to a very good approximation, the off-diagonal 
terms of the mass matrix in the $(\tilde B, {\tilde W}_{\scriptscriptstyle 3}, {\tilde H}_{1}^{\scriptscriptstyle 0},
{\tilde H}_{2}^{\scriptscriptstyle 0})$ basis are negligible as compared to 
$M_{\scriptscriptstyle 1}, M_{\scriptscriptstyle 2}$ and $\mu$. The physical mass eigenstates 
${\tilde {\chi}}_{i}^{\scriptscriptstyle 0}, (i=1,\ldots, 4)$ are predominantly 
$\tilde B, {\tilde W}_{\scriptscriptstyle 3}, 
({\tilde H}_{1}^{\scriptscriptstyle 0}+{\tilde H}_{2}^{\scriptscriptstyle 0})/\sqrt{2}$ and $({\tilde H}_{1}^{\scriptscriptstyle 0}-
{\tilde H}_{2}^{\scriptscriptstyle 0})/\sqrt{2}$, and their corresponding masses are close 
to $M_{\scriptscriptstyle 1}, M_{\scriptscriptstyle 2}, |\mu|$ and $|\mu|$ respectively.

\section{Effective action for the electroweak gauge bosons to one-loop in the large
SUSY masses limit.}

\hspace*{0.5cm} Our aim is to compute the effective action for the standard particles,
$\Gamma_{eff}[\phi]$, that is defined through functional integration of all the
sparticles of the MSSM. In short notation it is defined by,
\begin{equation}
\label{eq:gammaeff}
{\rm e}^{i\Gamma_{eff}[\phi]}=\int [{\rm d}\tilde\phi]\,{\rm e}^{i \Gamma_{\rm MSSM}
  [\phi,\tilde\phi]} \,\,,\hspace*{0.2cm}
\label{eq:gammaeffmssm}
\Gamma_{\rm MSSM}[\phi,\tilde\phi] \equiv \int \dx{\cal L}_{\rm
  MSSM}(\phi,\tilde \phi)\,\,;\,\,{\rm d}x\equiv{\rm d}^4x \,,
\end{equation}
where $\phi=l,q,A,W^\pm,Z,g,H$ are the SM particles; $\tilde \phi=\tilde l,\tilde q,\tilde A,
\w,\tilde Z,\tilde g,\tilde H$ their supersymmetric partners, and 
${\cal L}_{\rm MSSM}$ is the lagrangian of the MSSM~\cite{GEISHA}.

In this paper, we are interested in the part of the effective action that
contains the two point Green functions with external gauge bosons
$\Gamma_{\mu\nu}^{X\, Y} (k)$, $X, Y=A,Z,W^\pm$. This will allow us to study the 
decoupling properties of the gauge boson self-energies and from them we will deduce the 
corresponding analytical expressions for the well known parameters $S, T$ and $U$. The 
computation of the effective action has been  performed at the one loop level by using 
dimensional regularization and including the integration of all the
sfermions $\sfe$, the neutralinos $\sne$ and the charginos $\scp$. 

The effective action can be written as:
\begin{equation}
\label{eq:gammaeffV}
{\rm e}^{i\Gamma_{eff}[V]}=\int [{\rm d}\sfe]\, [{\rm d}\sfe^*]\, 
[{\rm d}\scp]\, [{\rm d}\bar{\sg}^+]\, [{\rm d}\sne]\,
{\rm e}^{i \Gamma_{\rm MSSM}[V,\sfe, \scp,\sne]}
\end{equation}
where $\tilde{f} = \tilde{q}, \tilde{l} \hspace*{0.3cm}; V=W^\pm, Z, A$ and:
\begin{equation}
\label{eq:gammaMSSM}
 \Gamma_{\rm MSSM}[V,\sfe, \scp,\sne] \equiv 
 \Gamma_0[V]+\Gamma_{\sfe}[V,\sfe]+\Gamma_{\sg}[V,\sg]+\Gamma_{\sg}[\sfe,\sg,f].
\end{equation}
In this formula $\Gamma_0[V]$ is the quadratic action for gauge bosons which is taken generically
in an arbitrary 
$R_\xi$ gauge, and $\Gamma_{\sfe}[V,\sfe]$ and $\Gamma_{\sg}[V,\sg]$ are the actions for the 
sfermions and the neutralinos and charginos respectively. Notice that the last term  
$\Gamma_{\sg}[\sfe,\sg,f]$, which includes the interactions among $\sfe, \sg$ and $f$, does 
not contribute to $\Gamma_{eff}[V]$ to one-loop. Therefore, the formula of the effective 
action (\ref{eq:gammaeffV}) can be factorized into three pieces: 
\begin{equation}
\begin{array}{l}
\displaystyle 
e^{i \Gamma_{eff} [{\scriptscriptstyle V}]} = e^{i \Gamma_{o} [{\scriptscriptstyle V}]} 
e^{i \Gamma_{eff}^{\tilde{f}} [{\scriptscriptstyle V}]}
e^{i \Gamma_{eff}^{\tilde{\chi}} [{\scriptscriptstyle V}]} 
\end{array}
\end{equation}
where,
\begin{equation}
\begin{array}{l}
\label{eq:gammaeffF}
\displaystyle 
e^{i \Gamma_{eff}^{\tilde{f}} [{\scriptscriptstyle V}]} = \int [d\tilde{f}] [d\tilde{f}^{*}] 
e^{i \Gamma_{\tilde{f}} [{\scriptscriptstyle V},\tilde{f}]} \,\,\,, \hspace*{0.3cm}
\label{eq:gammaeffCN}
\displaystyle e^{i \Gamma_{eff}^{\tilde{\chi}} [{\scriptscriptstyle V}]} = 
\int [d\tilde{\chi}^{+}] [d\bar{\tilde{\chi}}^{+}] [d\tilde{\chi}^{o}]  
e^{i \Gamma_{\tilde{\chi}} [{\scriptscriptstyle V},\tilde{\chi}]} 
\end{array}
\end{equation}
\hspace*{0.5cm} In order to perform the functional integration, it is convenient to write the
classical action in terms of operators. We computed $\Gamma_{eff}^{\tilde{f}} [V]$ and $\Gamma_{eff}^{\tilde{\chi}} [V]$ 
separately by using the standard path integral techniques. The details of the computation 
can be seen in \cite{GEISHA}.

If we keep just the terms that contribute to the two-point functions, we find that the 
effective action generated from sfermions and charginos and neutralinos integration 
can be written as, 
\begin{eqnarray}
\displaystyle
\label{eq:eff} 
\Gamma_{eff}^{\tilde{f}} [V] &=& i \Tr ({A_{\tilde{f}}^{(0)}}^{-1} A_{\tilde{f}}^{(2)}) - 
\frac{i}{2} \Tr ({A_{\tilde{f}}^{(0)}}^{-1} A_{\tilde{f}}^{(1)})^{2} + O(V^{3})\,\,, \\
\label{eq:efcn}
\Gamma_{eff}^{\tilde{\chi}} [V] &=& \frac{i}{2} \Tr ({A_{+}^{(o)}}^{-1} A_{+}^{(1)})^{2} + 
\frac{i}{4} \Tr ({A_{o}^{(o)}}^{-1} A_{o}^{(1)})^{2} + 
i \Tr ({A_{o}^{(o)}}^{-1} A_{o+}^{(1)} {A_{+}^{(o)}}^{-1} A_{+o}^{(1)})+O(V^{3})\,\,, 
\end{eqnarray}
where the operators are,
\begin{eqnarray}
  \label{eq:opersfdef}
  A_{\sfe xy}^{(0)}&\equiv&(-\Box-\tilde M_{f}^2)_x\delta_{xy}\,\,, \hspace*{0.5cm}
  A_{0\,xy}^{(0)}\equiv \left(i \slas{\partial}-\tilde M^0\right)_x \delta_{xy}\,\,, \hspace*{0.5cm}
  A_{+\,xy}^{(0)} \equiv \left(i \slas{\partial}-\tilde M^+\right)_x \delta_{xy}\nonumber\\
  A_{\sfe xy}^{(1)}&\equiv&\left[-ie\left(\partial_\mu A^\mu \hat{Q}_f+2\,\hat{Q}_f
    A_\mu\partial^\mu\right)-\frac{ig}{c_w}\left(\partial_\mu Z^\mu \hat{G}_f+2\,\hat{G}_f
    Z_\mu\partial^\mu\right)\right.\nonumber\\
  & &-\left.\frac{ig}{\sqrt{2}}\left(\partial_\mu W^{+\mu} \Sigma_f^{tb}+2\,
    \Sigma_f^{tb} W_\mu^+\partial^\mu\right)\right]_x\delta_{xy}+{\rm h.c.}\nonumber\\
  A_{\sfe xy}^{(2)}&\equiv&\left(e^2 \hat{Q}_f^2 A_\mu
  A^\mu+\frac{2\,g\,e}{c_w}A_\mu Z^\mu
  \hat{Q}_f\hat{G}_f+\frac{g^2}{c_w^2}\hat{G}_f^2 Z_\mu Z^\mu+\frac{1}{2}g^2
  \Sigma_f W_\mu^+ W^{\mu-}\right)_x \delta_{xy}\nonumber\\
  A_{0\,xy}^{(1)}&\equiv&\left[\frac{g}{c_w}Z_\mu \gamma^\mu\left(O_L^{\prime\prime}
  P_L+O_R^{\prime\prime} P_R\right)\right]_x \delta_{xy}\,\,, \hspace*{0.2cm} 
  A_{+\,xy}^{(1)} \equiv \left[\frac{g}{c_w}Z_\mu \gamma^\mu\left(O_L^{\prime}
  P_L+O_R^{\prime} P_R\right)-e\,A_\mu \gamma^\mu\right]_x \delta_{xy}\nonumber\\
  A_{0+\,xy}^{(1)}&\equiv&\left[g\, W_\mu^- \gamma^\mu\left(O_L
      P_L+O_R P_R\right)\right]_x \delta_{xy}\,\,, \hspace*{0.3cm} 
  A_{+0\,xy}^{(1)} \equiv \left[g\, W_\mu^+ \gamma^\mu\left(O_L^{+}
      P_L+O_R^{+} P_R\right)\right]_x \delta_{xy} \,\,.
\end{eqnarray}

In all these expressions and in the following, we use the compact notation \cite{GEISHA} where 
$\tilde{f}$ is a four-entries column vector including sfermions of all types and the 
sum $\sum_{\tilde{f}}$ is over the three generations and, in the case of squarks, it 
runs also over the $N_{c}$ color indexes. We have introduced as well two column 
vectors, $\tilde{\chi}^{o}$, with components $\tilde{\chi}^{o}_{i}$, $(i=1,2,3,4)$, and 
$\tilde{\chi}^{+}$, which components are $\tilde{\chi}^{+}_{i}$, $(i=1,2)$.
The corresponding mass matrices are:
\begin{equation}
\begin{array}{l}
\displaystyle
\tilde{M}_{f}^{2} = \diag(\tilde{m}_{t_{1}}^{2},\tilde{m}_{t_{2}}^{2},
\tilde{m}_{b_{1}}^{2},\tilde{m}_{b_{2}}^{2})
\hspace*{0.35cm} if \hspace*{0.35cm} \tilde{f} = \tilde{q}\,\,; \hspace*{0.3cm}
\tilde{M}_{f}^{2} = \diag(\tilde{m}_{\nu}^{2},0,\tilde{m}_{\tau_{1}}^{2},
\tilde{m}_{\tau_{2}}^{2})
\hspace*{0.35cm} if \hspace*{0.35cm} \tilde{f} = \tilde{l}\,\,,
\end{array}
\end{equation}
\begin{equation}
\begin{array}{l}
\displaystyle
\tilde{M}^{+} = \diag(\tilde{M}_{1}^{+}, \tilde{M}_{2}^{+})\,\,,\hspace*{0.5cm} 
\tilde{M}^{0} = \diag(\tilde{M}_{1}^{0}, \tilde{M}_{2}^{0}, \tilde{M}_{3}^{0}, 
\tilde{M}_{4}^{0})
\end{array}
\end{equation}
The coupling matrices $\hat{Q_{f}}, \hat{G_{f}}, \Sigma_{f}^{tb}, \Sigma_{f}$ and 
$O_{\scriptscriptstyle L,R}$, ${O'}_{\scriptscriptstyle L,R}$,
${O''}_{\scriptscriptstyle L,R}$ can be found in \cite{GEISHA}.

Notice that, diagrammatically, the first and second terms in eq.~(\ref{eq:eff}) give 
the two types of one-loop contributions with all kind of sfermions in the loop, the first
 term in eq.~(\ref{eq:efcn}) gives the 
one-loop contributions with charginos in the loop, the second term is the corresponding 
contribution with neutralinos in the loop and the last one gives the mixed one-loop 
contributions with both charginos and neutralinos in the loop. 

In order to get the explicit expressions for the two-point functions one must work 
out the traces in the above formulae. Basically one must substitute all the operators, 
express the one-loop integrals in momentum space of D dimensions, compute all the appearing 
Dirac traces and Fourier transform the result back to the position space. The traces also 
involve to perform the sum in the corresponding matrix indexes, the sum over the various 
types of sfermions and the sum in color indexes in the case of squarks. We have done this 
computation, in addition, by diagrammatical methods
and we have found the same results. Notice that they are exact to one-loop.
The result for the effective action is \cite{GEISHA}:
\begin{eqnarray}
\displaystyle \Gamma_{eff} [V] &=& 
\frac{1}{2} \int dx dy A_{x}^{\mu} \Gamma_{\mu \nu}^{AA} (x,y) A_{y}^{\nu} + 
\frac{1}{2} \int dx dy Z_{x}^{\mu} \Gamma_{\mu \nu}^{ZZ} (x,y) Z_{y}^{\nu} +
 \frac{1}{2} \int dx dy A_{x}^{\mu} \Gamma_{\mu \nu}^{AZ} (x,y) Z_{y}^{\nu} \nonumber \\ 
&+& (A \leftrightarrow Z)+ \frac{1}{2} \int dx dy W_{x}^{+ \mu} 
\Gamma_{\mu \nu}^{WW} (x,y) W_{y}^{- \nu} + (+ \leftrightarrow -) + O( V ^{3})
\end{eqnarray}
where $\Gamma_{\mu \nu} (x,y)$ are the two-point functions for the electroweak gauge bosons 
in position space. 

The results for these two-point functions in momentum space are as follows:
\begin{eqnarray}
\label{eq:gammaAA} 
\displaystyle \Gamma^{A\, A}_{\mu\, \nu}(k) &=& 
{\Gamma_0}^{A\, A}_{\mu\, \nu}(k)
+i e^{2} \sum_{\tilde{f}} \left\{ 2 \sum_{a} I_{o}(\tilde{m}_{f_{a}}^{2})
(\hat{Q}_{f}^{2})_{aa} g_{\mu \nu} - \sum_{ab} (\hat{Q}_{f})_{ab} 
(\hat{Q}_{f})_{ba} I_{f_{\mu \nu}}^{ab}(k, \tilde{m}_{f_{a}}, \tilde{m}_{f_{b}})
\right\} \nonumber \\
\displaystyle &+& 2 i e^{2} \sum_{i=1}^{2}  
\left\{ T_{\mu \nu}^{ii}(k, \tilde{M}_{i}^{+}, \tilde{M}_{i}^{+}) +
 2 {\hat{I}}^{ii}(k, \tilde{M}_{i}^{+}, \tilde{M}_{i}^{+}) g_{\mu \nu} \right\} \\ 
\nonumber \\
\label{eq:gammaZZ} 
\displaystyle \Gamma^{Z\, Z}_{\mu\, \nu} (k) &=& {\Gamma_0}^{Z\, Z}_{\mu\, \nu}(k)
+i \frac{g^{2}}{c_{{\scriptscriptstyle W}}^{2}} 
\sum_{\tilde{f}} \left\{ 2 \sum_{a} I_{o}(\tilde{m}_{f_{a}}^{2}) 
(\hat{G}_{f}^{2})_{aa} g_{\mu \nu} - \sum_{ab} (\hat{G}_{f})_{ab} 
(\hat{G}_{f})_{ba} I_{f_{\mu \nu}}^{ab}(k , \tilde{m}_{f_{a}}, \tilde{m}_{f_{b}})
\right\} \nonumber \\
\hspace*{1.0cm} &+& \displaystyle  \frac{i}{2} \frac{g^{2}}{c_{{\scriptscriptstyle W}}^{2}} \sum_{i,j=1}^{4} 
\left\{ ({O''}_{{\scriptscriptstyle L}}^{ij} {O''}_{{\scriptscriptstyle L}}^{ji} 
+ {O''}_{{\scriptscriptstyle R}}^{ij} {O''}_{{\scriptscriptstyle R}}^{ji})
T_{\mu \nu}^{ij}(k, \tilde{M}_{i}^{0}, \tilde{M}_{j}^{0}) + 
2 ({O''}_{{\scriptscriptstyle L}}^{ij} {O''}_{{\scriptscriptstyle R}}^{ji} +
{O''}_{{\scriptscriptstyle R}}^{ij} {O''}_{{\scriptscriptstyle L}}^{ji}) 
{\hat{I}}^{ij}(k, \tilde{M}_{i}^{0}, \tilde{M}_{j}^{0})  g_{\mu \nu} \right\} \nonumber \\
\hspace{1.0cm} \displaystyle &+& i \frac{g^{2}}{c_{{\scriptscriptstyle W}}^{2}} \sum_{i,j=1}^{2} 
\left\{ ({O'}_{{\scriptscriptstyle L}}^{ij} {O'}_{{\scriptscriptstyle L}}^{ji} + 
{O'}_{{\scriptscriptstyle R}}^{ij} {O'}_{{\scriptscriptstyle R}}^{ji})
T_{\mu \nu}^{ij}(k,  \tilde{M}_{i}^{+}, \tilde{M}_{j}^{+}) + 
2 ({O'}_{{\scriptscriptstyle L}}^{ij} {O'}_{{\scriptscriptstyle R}}^{ji} +
{O'}_{{\scriptscriptstyle R}}^{ij} {O'}_{{\scriptscriptstyle L}}^{ji}) 
{\hat{I}}^{ij}(k,  \tilde{M}_{i}^{+}, \tilde{M}_{j}^{+}) g_{\mu \nu}  \right\} \nonumber \\
\\ \nonumber
\label{eq:gammaAZ} 
\displaystyle \Gamma^{A\, Z}_{\mu\, \nu}(k) &=&\Gamma^{Z\, A}_{\mu\, \nu}(k)= 
\frac{i g e}{\cw} \sum_{\widetilde{f}} \left\{ 
2 \sum_{a} I_0(\tilde{m}^2_{f_a}) (\widehat{Q}_{f} \widehat{G}_{f})_{a\, a}  g_{\mu\, \nu}
- \sum_{ab}  (\widehat{Q}_{f})_{a\, b} (\widehat{G}_{f})_{b\, a} 
I^{a\,b}_{f_{\mu\, \nu}}(k , \tilde{m}_{f_{a}}, \tilde{m}_{f_{b}}) \right\} \nonumber \\
\hspace*{1cm} \displaystyle &-& \frac{i g e}{\cw}  \sum_{i=1}^2  
\left(  {O'}^{i\, i}_L + {O'}^{i\, i}_R \right)\left(
T_{\mu\, \nu}^{i\, i}(k, \tilde{M}_{i}^{+}, \tilde{M}_{i}^{+}) 
+ 2 {\hat{I}}^{ii}(k, \tilde{M}_{i}^{+}, \tilde{M}_{i}^{+}) g_{\mu\, \nu}\right)
\\
\label{eq:gammaWW} 
\displaystyle \Gamma^{W\, W}_{\mu\, \nu}(k) &=&
{\Gamma_0}^{W\, W}_{\mu\, \nu}(k)
+ \frac{i g^2}{2} \sum_{\widetilde{f}}  \left\{ \sum_{a} (\Sigma_{f})_{a\, a}
I_0(\tilde{m}^2_{f_a}) g_{\mu\, \nu}-\sum_{a,b} (\Sigma_{f}^{t\,b})_{a\, b} 
(\Sigma_{f}^{t\,b})_{a\, b} I^{a\,b}_{f_{\mu\, \nu}}(k, \tilde{m}_{f_{a}}, \tilde{m}_{f_{b}}) 
\right\} \nonumber \\
\displaystyle &+& i g^{2} \sum_{i=1}^4 \sum_{j=1}^2 \left\{
\left({O}^{i\, j }_L {O}^{+\, j\, i}_L + {O}^{i\, j}_R  {O}^{+\, j\, i}_R \right)
T^{i\,j}_{\mu\, \nu}(k, \tilde{M}_{i}^{0}, \tilde{M}_{j}^{+}) \right. \nonumber \\
\displaystyle  &+& \left.2 \left({O}^{i\,j}_L {O}^{+\,j\,i}_R + {O}^{i\, j}_R  {O}^{+\, j\, i}_L \right)
{\hat{I}}^{ij}(k, \tilde{M}_{i}^{0}, \tilde{M}_{j}^{+})  g_{\mu\, \nu} \right\} 
\end{eqnarray}
Here the indexes $a$ and $b$ run from one to four, corresponding to the four entries of the
column vector $\tilde{f}$. The indexes $i, j$ vary as $i,j=1,2,3,4$ if they refer to neutralinos
 and as $i,j=1,2$ if 
they refer to charginos. ${\Gamma_0}^{V\, V}_{\mu\, \nu}$ $(V = Z, W)$ and 
${\Gamma_0}^{A\, A}_{\mu\, \nu}$ are the two-point functions at tree level, which are defined by:
\begin{equation}
{\Gamma_0}^{V\, V}_{\mu\, \nu} (k) = (M_{\scriptscriptstyle V}-k^{2}) g_{\mu\, \nu} + 
\left(1 - \frac{1}{\xi_{{\scriptscriptstyle V}}}\right) k_{\mu} k_{\nu} \,\,;
\,\, {\Gamma_0}^{A\, A}_{\mu\, \nu} = -k^{2}g_{\mu\, \nu} + 
\left(1 - \frac{1}{\xi_{{\scriptscriptstyle A}}}\right) k_{\mu} k_{\nu} \,\,, 
\end{equation}
and the one-loop integrals $I_{0}(\mfa)\,, I^{a\,b}_{f_{\mu\, \nu}}(k, \tilde{m}_{f_{a}}, 
\tilde{m}_{f_{b}})\,, T^{i\,j}_{\mu\, \nu}(k, \tilde{M}_{i}, \tilde{M}_{j})\,, 
{\hat I}^{i\,j}(k, \tilde{M}_{i}, \tilde{M}_{j})$ are defined in dimensional regularization by,
\begin{equation}
\label{eq:int0}
I_{0}(\mfa) =  \int d\widehat{q} \frac{1}{\left[q^2 - \mfa \right]} \,\,;\hspace*{0.3cm}
I^{a\,b}_{f_{\mu\, \nu}}(k, \tilde{m}_{f_{a}}, \tilde{m}_{f_{b}}) =  \int d\widehat{q} 
\frac{(2q+k)_{\mu} (2q+k)_{\nu}}
{\left[(k+q)^2 - \tilde{m}^2_{f_a}\right] \left[q^2 - \tilde{m}^2_{f_b}\right]}
\end{equation}
\begin{equation}
T^{i\,j}_{\mu\, \nu}(k, \tilde{M}_{i}, \tilde{M}_{j})= \int d\widehat{q}
\frac{(4q_{\mu}q_{\nu}-2g_{\mu\, \nu}g^{\alpha\, \beta}q_{\alpha}q_{\beta}+
2(q_{\mu}k_{\nu}+q_{\nu}k_{\mu})-2g_{\mu\, \nu}g^{\alpha\, \beta}q_{\alpha}k_{\beta})}
{\left[(k+q)^2 - \tilde{m}^2_{f_a}\right] \left[q^2 - \tilde{m}^2_{f_b}\right]}
\end{equation}
\begin{equation}
\label{eq:intM}
{\hat I}^{i\,j}(k, \tilde{M}_{i}, \tilde{M}_{j}) = \int d\widehat{q} 
\frac{\tilde{M}_{i} \tilde{M}_{j} }
{ \left[q^2 - \tilde{M}^2_{i}\right]\left[(k+q)^2 - \tilde{M}^2_{j}\right]}
\end{equation}
\hspace*{0.5cm}Since we are interested in the large mass limit of the SUSY particles we need to have at 
hand not just the exact results of the above mentioned integrals but their asymptotic expressions to
be valid in that limit. We have analized the integrals by means of the so-called m-Theorem 
\cite{GMR}. This theorem provides a powerful technique to study the asymptotic behaviour of Feynman 
integrals in the limit where some of the masses are large. Notice that this is not
trivial since some of these integrals are divergent and the interchange of the 
integral with the limit is not allowed. Thus, one should first compute the integrals in 
dimensional regularization and at the end take the large mass limit. Instead of this direct
way it is also possible to proceed as follows: First, in order to decrease the ultraviolet
divergent degree, one rearranges the integrand through
algebraic manipulations up to separate the Feynman integral into a divergent part, which 
can be evaluated exactly using the standard techniques of dimensional regularization, and a convergent part that satisfies
the requirements demanded by the m-Theorem and therefore, goes to zero in the infinite mass
limit. By means of this procedure the correct asymptotic behaviour
of the integrals is guaranteed. Some examples of the computation of the Feynman integrals by means of the m-Theorem as well
as details of this theorem are given in \cite{GEISHA}. The results for the above one loop integrals
in the large masses limit are as follows:
\begin{eqnarray}
\label{eq:io}
\displaystyle I_{0}(\mfa) &=& \frac{i}{16 \pi^{2}} 
\left( {\Delta}_\epsilon+1-\log \frac{\mfa}{\mu_{o}^{2}} \right) {\mfa}\nonumber \\
\displaystyle I^{ab}_{{\scriptscriptstyle f}} (k, \tilde{m}_{f_{a}}, \tilde{m}_{f_{b}})
&=&  \frac{i}{16 \pi^{2}} \left\{
(\mfa+\mfb) \left({\Delta}_\epsilon+1-\log \frac{(\mfa+\mfb)}{2\mu_{o}^{2}}\right)g_{\mu\, \nu}-
\frac{1}{3} k^{2}  \left({\Delta}_\epsilon-\log 
\frac{(\mfa+\mfb)}{2\mu_{o}^{2}}\right)g_{\mu\, \nu} \right. \nonumber \\
\nonumber \\
&&+ \frac{1}{3}\left. k_{\mu}k_{\nu}\left({\Delta}_\epsilon-
\log \frac{(\mfa+\mfb)}{2\mu_{o}^{2}}\right)\right\}\nonumber \\
\nonumber \\
\displaystyle T^{ij} (k, \tilde{M}_{i}, \tilde{M}_{j}) &=& \frac{i}{16 \pi^{2}} 
\left\{-(\mi+\mj) \left( {\Delta}_\epsilon-\log \frac{(\mi+\mj)}{2\mu_{o}^{2}}\right) +
\frac{2}{3} k^{2} \left({\Delta}_\epsilon-\frac{1}{2} -\log 
\frac{(\mi+\mj)}{2\mu_{o}^{2}}\right)g_{\mu\, \nu} \right. \nonumber 
\\
&&- \frac{2}{3}k_{\mu}k_{\nu}\left. \left( {\Delta}_\epsilon-
\log \frac{(\mi+\mj)}{2\mu_{o}^{2}} \right)\right\}\,,\nonumber \\
\label{eq:intf}
\displaystyle {\hat I}^{i\,j} (k, \tilde{M}_{i}, \tilde{M}_{j}) &=& 
\frac{i}{16 \pi^{2}} \left\{ \frac{1}{2} (\mi+\mj) \left({\Delta}_\epsilon-
\log \frac{(\mi+\mj)}{2\mu_{o}^{2}}\right)+ \frac{1}{6} k^{2} \right. \nonumber\\
&&
\displaystyle
- \left. \frac{1}{2} (\tilde{M}_{i}-\tilde{M}_{j})^{2} \left({\Delta}_\epsilon-\log \frac{(\mi+\mj)}{2\mu_{o}^{2}}
\right) \right\}\,.
\end{eqnarray}
where $\mu_{o}$ is the usual mass scale of dimensional regularization and,
\begin{equation}
\hspace*{0.6cm} \displaystyle {\Delta}_\epsilon=\frac{2}{\epsilon }-{\gamma }_{\epsilon} 
+\log (4\pi) \hspace*{0.2cm}, \hspace*{0.2cm} \epsilon = 4-D\,.
\end{equation}
\hspace*{0.5cm}Finally, we define the self-energies $\Sigma^{X\, Y} (k)$ and the $R^{X\, Y} (k)$ functions, 
from the two-point functions as usual,\\
\begin{equation}
\label{eq:gasi}
\Gamma^{X\, Y}_{\mu\, \nu} (k) = {\Gamma_0}^{X\, Y}_{\mu\, \nu} (k) +
\Sigma^{X\, Y} (k)  g_{\mu\, \nu} + R^{X\, Y} (k) k_{\mu} k_{\nu}\,\,.
\end{equation}

The asymptotic expressions for the $\Sigma^{{\scriptscriptstyle XY}}$ and 
$R^{{\scriptscriptstyle XY}}$ functions, in the large sparticle masses limit and for each 
sector, can be obtained from our results of eqs.(\ref{eq:gammaAA}-\ref{eq:gammaWW}) and by using 
the formulae of eqs.(\ref{eq:intf}). We find the following results:
\newpage
\subsection{Squarks sector:}
For $\hspace*{0.3cm}\tilde{m}_{t_{1}}^{2}, \tilde{m}_{t_{2}}^{2}, 
\tilde{m}_{b_{1}}^{2}, \tilde{m}_{b_{2}}^{2} \gg k^{2}\,\,\,, 
\hspace*{0.2cm}|\tilde{m}_{t_{1}}^{2} - \tilde{m}_{t_{2}}^{2}| \ll
|\tilde{m}_{t_{1}}^{2} + \tilde{m}_{t_{2}}^{2}| \hspace*{0.1cm};\hspace*{0.2cm}
|\tilde{m}_{b_{1}}^{2} - \tilde{m}_{b_{2}}^{2}| \ll 
|\tilde{m}_{b_{1}}^{2} + \tilde{m}_{b_{2}}^{2}| \,\,\,,$ and \\ \\$\hspace*{0.1cm}
|\tilde{m}_{t_{i}}^{2} - \tilde{m}_{b_{j}}^{2}| \ll
|\tilde{m}_{t_{i}}^{2} + \tilde{m}_{b_{j}}^{2}| \hspace*{0.2cm} (i,j=1,2) \,,$ we get:
\begin{eqnarray}
\label{eq:SumqAA}
\displaystyle \Sigma_{\tilde{q}}^{{\scriptscriptstyle AA}} (k) &=& 
-N_{c} \frac{e^{2}}{16 \pi^{2}} k^{2} \frac{1}{3} \sum_{\tilde{q}}
\left\{\frac{10}{9} \Delta_{\epsilon} -\frac{4}{9} \left(\log \frac{\tilde{m}_{t_{1}}^{2}}{\mu_{o}^{2}} 
+\log \frac{\tilde{m}_{t_{2}}^{2}}{\mu_{o}^{2}}\right) -\frac{1}{9} \left(  
\log \frac{\tilde{m}_{b_{1}}^{2}}{\mu_{o}^{2}}+\log \frac{\tilde{m}_{b_{2}}^{2}}{\mu_{o}^{2}}
\right) \right\}, \nonumber \\
\\
\label{eq:SumqAZ}
\displaystyle \Sigma_{\tilde{q}}^{{\scriptscriptstyle AZ}} (k) &=& 
-N_{c} \frac{e^{2}}{16 \pi^{2}} k^{2} \frac{1}
{3 s_{{\scriptscriptstyle W}} c_{{\scriptscriptstyle W}}} 
\displaystyle \sum_{\tilde{q}} \left\{ \left(\frac{1}{2}-\frac{10}{9}s_{{\scriptscriptstyle W}}^{2}\right)
\Delta_{\epsilon} - 
\frac{2}{3} \left(\frac{1}{2} c_{t}^{2} - \frac{2}{3} s_{{\scriptscriptstyle W}}^{2}\right)
\log \frac{\tilde{m}_{t_{1}}^{2}}{\mu_{o}^{2}} 
\right.\nonumber \\
\displaystyle &-& \frac{2}{3} \left(\frac{1}{2} s_{t}^{2} - 
\frac{2}{3} s_{{\scriptscriptstyle W}}^{2}\right) 
\log \frac{\tilde{m}_{t_{2}}^{2}}{\mu_{o}^{2}}+
\frac{1}{3} \left(-\frac{1}{2} c_{b}^{2} + \frac{1}{3} s_{{\scriptscriptstyle W}}^{2}\right) 
\log \frac{\tilde{m}_{b_{1}}^{2}}{\mu_{o}^{2}}
+ \left.\frac{1}{3} 
\left(-\frac{1}{2} s_{b}^{2} + \frac{1}{3} s_{{\scriptscriptstyle W}}^{2}\right) 
\log \frac{\tilde{m}_{b_{2}}^{2}}{\mu_{o}^{2}} \right\}, \nonumber \\
\\
\label{eq:SumqZZ}
\displaystyle \Sigma^{{\scriptscriptstyle ZZ}}_{\tilde{q}} (k) &=& -N_{c} 
\frac{e^{2}}{16 \pi^{2}} \frac{1}
{s_{{\scriptscriptstyle W}}^{2} c_{{\scriptscriptstyle W}}^{2}} \sum_{\tilde{q}} \left\{ \left[
-\frac{1}{2} c_{t}^{2} s_{t}^{2} h({\tilde{m}}_{t_{1}}^{2}, {\tilde{m}}_{t_{2}}^{2}) 
- \frac{1}{2} c_{b}^{2} s_{b}^{2} h({\tilde{m}}_{b_{1}}^{2}, 
{\tilde{m}}_{b_{2}}^{2}) \right. \right] \nonumber \\
\displaystyle &+& \frac{1}{3} k^{2} \left[
\left(\frac{c_{t}^{2}}{2} - \frac{2 s_{{\scriptscriptstyle W}}^{2}}{3}\right)^{2} 
\left(\Delta_\epsilon - \log \frac{\tilde{m}_{t_{1}}^{2}}{\mu_{o}^{2}}\right) \right. 
+\left(\frac{s_{t}^{2}}{2} - \frac{2 s_{{\scriptscriptstyle W}}^{2}}{3}\right)^{2} 
\left(\Delta_\epsilon - \log \frac{\tilde{m}_{t_{2}}^{2}}{\mu_{o}^{2}}\right) \nonumber \\
 &+& \displaystyle
\left(-\frac{c_{b}^{2}}{2} + \frac{s_{{\scriptscriptstyle W}}^{2}}{3}\right)^{2}
\left(\Delta_\epsilon - \log \frac{\tilde{m}_{b_{1}}^{2}}{\mu_{o}^{2}}\right) + 
\left(-\frac{s_{b}^{2}}{2} + \frac{s_{{\scriptscriptstyle W}}^{2}}{3}\right)^{2}
\left(\Delta_\epsilon - \log \frac{\tilde{m}_{b_{2}}^{2}}{\mu_{o}^{2}}\right) \nonumber \\
 &+& \displaystyle \left. \left.\frac{1}{2} s_{t}^{2} c_{t}^{2} 
\left(\Delta_\epsilon - \log \frac{\tilde{m}_{t_{1}}^{2} + \tilde{m}_{t_{2}}^{2}}
{2 \mu_{o}^{2}}\right) + \frac{1}{2} s_{b}^{2} c_{b}^{2} 
\left(\Delta_\epsilon - \log \frac{\tilde{m}_{b_{1}}^{2} + \tilde{m}_{b_{2}}^{2}}{2 
\mu_{o}^{2}}\right) \right] \right\}, \\
\nonumber \\
\label{eq:SumqWW}
\displaystyle \Sigma_{\tilde{q}}^{{\scriptscriptstyle WW}} (k) &=& 
-N_{c} \frac{e^{2}}{16 \pi^{2}} \frac{1}{s_{{\scriptscriptstyle W}}^{2}} 
\sum_{\tilde{q}} \left\{ \left[ 
-\frac{1}{2} c_{t}^{2} c_{b}^{2} h (\tilde{m}_{t_{1}}^{2}, 
\tilde{m}_{b_{1}}^{2}) -\frac{1}{2} c_{t}^{2} s_{b}^{2} 
h ({\tilde{m}}_{t_{1}}^{2}, {\tilde{m}}_{b_{2}}^{2}) \right.\right. \nonumber \\
&-& \displaystyle \frac{1}{2} s_{t}^{2} c_{b}^{2} h (\tilde{m}_{t_{2}}^{2}, 
\tilde{m}_{b_{1}}^{2}) \left. -\frac{1}{2} s_{t}^{2} s_{b}^{2} 
h (\tilde{m}_{t_{2}}^{2}, \tilde{m}_{b_{2}}^{2}) \right]
+\frac{1}{6} k^{2} \left[ 
\Delta_{\epsilon} -c_{t}^{2} c_{b}^{2} 
\log \frac{\tilde{m}_{t_{1}}^{2} + \tilde{m}_{b_{1}}^{2}}{2 \mu_{o}^{2}}
\right. \nonumber \\
\displaystyle &-& c_{t}^{2} s_{b}^{2} 
\log \frac{\tilde{m}_{t_{1}}^{2} + \tilde{m}_{b_{2}}^{2}}
{2 \mu_{o}^{2}} - s_{t}^{2} c_{b}^{2} 
\log \frac{\tilde{m}_{t_{2}}^{2} + \tilde{m}_{b_{1}}^{2}}
{2 \mu_{o}^{2}} -\left. \left.  s_{t}^{2} s_{b}^{2} 
\log \frac{\tilde{m}_{t_{2}}^{2} + \tilde{m}_{b_{2}}^{2}}
{2 \mu_{o}^{2}}\right] \right\}\,\,, 
\end{eqnarray}
where $\sw^{2}=sin^{2}{\theta}_{\scriptscriptstyle W}$ and $c_{f}=cos\theta_{f}, 
s_{f}=sin\theta_{f}$, with $\theta_{f}$ being the mixing angle in the $f$-sector.

The function $h({m}_{1}^{2}, {m}_{2}^{2})$ has been defined in \cite{GEISHA} and its 
asymptotic behaviour for large masses $\,{m}_{1}^{2}\,\,, {m}_{2}^{2}\,$ and 
$\,\,|{m}_{1}^{2} - {m}_{2}^{2}| \ll |{m}_{1}^{2} + {m}_{2}^{2}|$ is given by:
\begin{equation}
h({m}_{1}^{2}, {m}_{2}^{2}) \longrightarrow
\frac{{m}_{1}^{2} - {m}_{2}^{2}}{2} \left[
\frac{({m}_{1}^{2} - {m}_{2}^{2})}{({m}_{1}^{2} + {m}_{2}^{2})} +
O{\left(\frac{{m}_{1}^{2} - {m}_{2}^{2}}{{m}_{1}^{2} + {m}_{2}^{2}}\right)}^{2}\right]
\end{equation}
\hspace*{0.5cm}For the $R^{{\scriptscriptstyle XY}} (k)$ functions we find expressions which
are similar to the part proportional to $k^{2}$ in $\Sigma^{{\scriptscriptstyle XY}} (k)$  
and with opposite sign. We omitt to write the explicit formulae here for brevity. They can, 
generically, be written as:
\begin{equation}
\label{eq:trampa}
R^{{\scriptscriptstyle XY}} (k) = -\left[ {\rm term}\hspace*{0.1cm}{\rm in} 
\hspace*{0.1cm} k^{2}\hspace*{0.1cm} {\rm of}\hspace*{0.1cm}
\Sigma^{{\scriptscriptstyle XY}} (k) \right] / k^{2}
\end{equation} 
\hspace*{0.5cm}As can be seen from eqs.(\ref{eq:SumqAA}-\ref{eq:SumqWW}), the asymptotic results in the 
large SUSY masses limit are of the generic form: $\Sigma^{X\, Y} (k)=\Sigma^{X\, Y}_{(0)}+
\Sigma^{X\, Y}_{(1)} k^{2}$ and $R^{X\, Y} (k)=R^{X\, Y}_{(0)}$, where $\Sigma^{X\, Y}_{(0)}, \Sigma^{X\, Y}_{(1)}$
and $R^{X\, Y}_{(0)}$ are functions of the large SUSY masses but are $k$ independent. As can
be easily shown, it implies that all the remaining dependence on the SUSY masses can be 
absorbed into a redefinition of the SM relevant parameters, $m_{\scriptscriptstyle W},
m_{\scriptscriptstyle Z}$ and $e$ and the gauge bosons wave functions. Therefore, the
decoupling of squarks in the two point functions does indeed occur.

Similar results have been obtained in the sleptons sector.

\subsection{Charginos and neutralinos sector:}
For $\hspace*{0.3cm} \tilde{M}_{1}^+{}^{2}, \tilde{M}_{2}^+{}^{2} \gg k^{2} \,\,\,, 
\hspace*{0.2cm} |\tilde{M}_{1}^+{}^{2} - \tilde{M}_{2}^+{}^{2}| \ll 
|\tilde{M}_{1}^+{}^{2} + \tilde{M}_{2}^+{}^{2}| \hspace*{0.1cm};\hspace*{0.2cm}
|\tilde{M}_{i}^o{}^{2} - \tilde{M}_{j}^o{}^{2}| \ll 
|\tilde{M}_{i}^o{}^{2} + \tilde{M}_{j}^o{}^{2}| \hspace*{0.2cm} (i,j=1,2,3,4)  \,\,\,,$ 
and \\ \\$\hspace*{0.1cm}
|\tilde{M}_{i}^o{}^{2} - \tilde{M}_{j}^+{}^{2}| \ll
|\tilde{M}_{i}^o{}^{2} + \tilde{M}_{j}^+{}^{2}| ,\hspace*{0.1cm} (i=1,2,3,4); 
\hspace*{0.1cm} (j=1,2) \,,$ we find:
\begin{eqnarray}
\label{eq:SumcnAA}
\displaystyle \Sigma_{\tilde{\chi}}^{{\scriptscriptstyle AA}} (k) &=& 
-\frac{e^{2}}{16 \pi^{2}} \frac{4}{3} k^{2} 
\left( 2 \Delta_{\epsilon}-\log \frac{\tilde{M}_{1}^{+^{2}}}{\mu_{o}^{2}}
-\log \frac{\tilde{M}_{2}^{+^{2}}}{\mu_{o}^{2}}\right) \\
\label{eq:SumcnAZ}
\displaystyle \Sigma_{\tilde{\chi}}^{{\scriptscriptstyle AZ}} (k) &=& 
\frac{e^{2}}{16 \pi^{2}}\frac{1}{s_{\scriptscriptstyle W} c_{\scriptscriptstyle W}} \frac{4}{3}k^{2}
\left\{ \left({\sw}^{2}-1\right)  \left(\Delta_{\epsilon}  
-\log \frac{\tilde{M}_{1}^+{}^{2}}{\mu_{o}^{2}}\right) 
+ \left({\sw}^{2}-\frac{1}{2}\right)  \left(\Delta_{\epsilon} 
-\log \frac{\tilde{M}_{2}^+{}^{2}}{\mu_{o}^{2}}\right) \right\}, \nonumber \\
\\ \nonumber
\label{eq:SumcnZZ}
\displaystyle \Sigma^{{\scriptscriptstyle ZZ}}_{\tilde{\chi}} (k) &=& 
-\frac{e^{2}}{16 \pi^{2}} \frac{1}
{s_{{\scriptscriptstyle W}}^{2} c_{{\scriptscriptstyle W}}^{2}} \left\{
-\frac{1}{2} {(\tilde{M}_{3}^{o} - \tilde{M}_{4}^{o})}^{2}
\left(\Delta_{\epsilon} - \log \frac{\tilde{M}_{3}^o{}^{2} 
+ \tilde{M}_{4}^o{}^{2}}{2 \mu_{o}^{2}}\right) \right. \nonumber \\
\displaystyle &+& \frac{4}{3} k^{2} \left[ \left(\sw^{2}-1\right)^{2}  
\left(\Delta_{\epsilon} - \log \frac{\tilde{M}_{1}^+{}^{2}}
{\mu_{o}^{2}}\right) + \left(\sw^{2}-\frac{1}{2}\right)^{2} 
\left(\Delta_{\epsilon} - \log \frac{\tilde{M}_{2}^+{}^{2}}{\mu_{o}^{2}}\right) \right.\nonumber \\
\displaystyle &+& \left.\left.\frac{1}{4}\left({\Delta}_{\epsilon} - 
\log \frac{\tilde{M}_{3}^o{}^{2}+\tilde{M}_{4}^o{}^{2}}{2 \mu_{o}^{2}}\right)
\right] \right\}, \\
\label{eq:SumcnWW}
\displaystyle \Sigma^{{\scriptscriptstyle WW}}_{\tilde{\chi}} (k) &=& 
-\frac{e^{2}}{16 \pi^{2}}  \frac{1}{s_{{\scriptscriptstyle W}}^{2}} 
\left\{ - 2 (\tilde{M}_{1}^{+} - \tilde{M}_{2}^{o})^{2} 
\left(\Delta_{\epsilon} - \log \frac{\tilde{M}_{1}^+{}^{2} + 
\tilde{M}_{2}^o{}^{2}}{2 \mu_{o}^{2}}\right) \right. \nonumber \\
\displaystyle &-& \frac{1}{2} (\tilde{M}_{2}^{+} - \tilde{M}_{3}^{o})^{2} 
\left(\Delta_{\epsilon} - \log \frac{\tilde{M}_{2}^+{}^{2} + 
\tilde{M}_{3}^o{}^{2}}{2 \mu_{o}^{2}}\right)-\frac{1}{2} (\tilde{M}_{2}^{+}-\tilde{M}_{4}^{o})^{2} 
\left(\Delta_{\epsilon} - \log \frac{\tilde{M}_{2}^+{}^{2} + 
\tilde{M}_{4}^o{}^{2}}{2 \mu_{o}^{2}}\right) \nonumber \\
\displaystyle &+& k^{2} \left[2\Delta_{\epsilon}- \frac{4}{3}\log \frac{\tilde{M}_{1}^+{}^{2} + 
\tilde{M}_{2}^o{}^{2}}{2 \mu_{o}^{2}}
-\frac{1}{3} \log \frac{\tilde{M}_{2}^+{}^{2} + 
\tilde{M}_{3}^o{}^{2}}{2 \mu_{o}^{2}} \left.
-\frac{1}{3} \log \frac{\tilde{M}_{2}^+{}^{2} + 
\tilde{M}_{4}^o{}^{2}}{2 \mu_{o}^{2}}\right] \right\}, 
\end{eqnarray}
\hspace*{0.5cm}Similarly to the squarks sector, the results for the 
$R^{{\scriptscriptstyle XY}} (k)$ functions, can be written generically as in 
eq.~(\ref{eq:trampa}).

If we consider the large ino masses limit which, as we have said in the above section, implies 
${\tilde{M}}_{1}^{+^{2}}\rightarrow M_{2}^{2}, \hspace*{0.1cm}
{\tilde{M}}_{2}^{+^{2}}\rightarrow\mu^{2}, \hspace*{0.1cm}{\tilde{M}}_{1}^{o^{2}}
\rightarrow M_{1}^{2}, \hspace*{0.1cm}{\tilde{M}}_{2}^{o^{2}}\rightarrow M_{2}^{2}$
and ${\tilde{M}}_{3}^{o^{2}}={\tilde{M}}_{4}^{o^{2}} \rightarrow \mu^{2}$, it is easy to see
that the remaining dependence on the SUSY masses is logarithmic. The 
$\Sigma^{{\scriptscriptstyle XY}} (k)$ functions are proportional
to $k^{2}$, and the $R^{{\scriptscriptstyle XY}} (k)$ functions are $k$ independent.
Therefore, similarly to the squarks sector, the decoupling of inos takes place.

In summary, from our results it is
clear that there is indeed decoupling in the two point electroweak gauge boson functions:
All SUSY effects can be absorbed into redefinitions of $m_{\scriptscriptstyle Z},
m_{\scriptscriptstyle W}, e$ and the wave functions of the gauge bosons $W^{\pm}, Z, A$, or
else they are suppressed by inverse powers of the heavy SUSY particles masses.

In addition, some comments can be done. First, these asymptotic expressions 
are completely general and depend just on the
physical masses of the SUSY particles and on the generic
coefficients $\hspace*{0.1cm}c_{q}\,, s_{q}\,, c_{l}\,, s_{l}\,, 
\hspace*{0.1cm}{O}^{j\, i}_{L,R}\,,$ 
$\hspace*{0.1cm} {O'}^{j\, i}_{L,R}\,, {O''}^{j\, i}_{L,R}$. Notice that they do 
not depent on the particular mechanism that generates the SUSY masses.

Second, it is worth to mention that in order to get the above asymptotic expressions 
not all the SUSY masses need to be compared with each other, but just the ones appearing 
in the same one-loop 
diagram. Thus, for instance, the self-energies $\Sigma^{A\, A}$ and $\Sigma^{A\, Z}$, where no 
mixed diagrams with different sfermions contribute, do not need of any reference on 
the relative size of the sfermion masses. $\Sigma^{Z\, Z}$ and $\Sigma^{W\, W}$, 
on the contrary, do require this comparison. In the case of $\Sigma^{Z\, Z}$ one 
needs to compare squarks of the same charge, sleptons of the same charge, charginos of 
the same charge and neutralinos among them. No comparison among sfermions of
different generations is required since we have not considered intergenerational
mixing in this work. In the case of $\Sigma^{W\, W}$ one needs to compare, in each
generation, the squarks of different charge, the sleptons of different charge and the 
netralinos with the charginos.The realistic and more interesting situation will be when all the
sparticles masses must be compared at the same time and, obviously, the final result 
will depend on the kind of SUSY hierarchy masses that had been previously established.
This will happen in the observables as $S, T$ and $U$ where all the four self-energies do contribute.
 
\section{Decoupling of sparticles in $S, T$ and $U$}
\label{sec:stu}
\renewcommand\baselinestretch{1.3}

\hspace*{0.5cm} The radiative corrections from SUSY particles to the observables $S, T$ and
$U$ have been analyzed exhaustively in the literature~\cite{CHA,HA2}, but neither their 
complete analytical expressions in the large sparticle masses limit  nor a general and 
systematic study of sparticles decoupling have been provided so far. We have obtained the
results to one loop for the analytical expressions of $S, T$ and $U$ in a
complete general form and analyzed under which particular conditions the sparticles 
decoupling takes place, discussing how and why does it occur in
the very special case of the MSSM with soft SUSY breaking terms~\cite{GEISHA}.

The definition that we use for $S, T$ and $U$ are the usual ones~\cite{PT}.
The contribution to $S, T$ and $U$ are known to be finite and well defined
separately for each sparticle sector, so that we can analyze them separately as
well. As we have already said we consider in this paper all the sparticle
contributions except that of the Higgs sector. 

The results for $S, T$ and $U$ can be obtained easily by using the corresponding 
expressions for the self-energies given eqs. (\ref{eq:SumqAA}-\ref{eq:SumcnWW}). 
Notice that, for each case, we must consider the corresponding conditions on the
masses above mentioned. Although the three parameters, $S, T$ and $U$ do not require the same set 
of conditions on the sparticle masses, the physical and realistic situation 
corresponds to have fixed all the SUSY spectra 
at once, and therefore all these conditions must hold together. 
\subsection{Squarks sector:}
\hspace*{0.5cm}By considering the conditions given at the beginning of section 3.1
together, is equivalent to say that all the squarks of the same generation have masses of 
similar large size. Interestingly, if we look just at the $S_{\tilde{q}}$ parameter,
there is apparently no decoupling since the dominant contribution goes as~\cite{GEISHA}:
\begin{equation}
\displaystyle 
S_{\tilde{q}} \rightarrow -\sum_{\tilde{q}} \frac{N_{c}}{36 \pi} 
\log \frac{\tilde{m}_{t_{1}}^{2}}{\tilde{m}_{b_{1}}^{2}},
\hspace{2.0cm} (\tilde{m}_{q_{i}}^{2} \gg k^{2})
\end{equation}
which under the corresponding conditions $|\tilde{m}_{t_{1}}^{2}-\tilde{m}_{t_{2}}^{2}| \ll 
|\tilde{m}_{t_{1}}^{2}+\tilde{m}_{t_{2}}^{2}|$ and 
$|\tilde{m}_{b_{1}}^{2}-\tilde{m}_{b_{2}}^{2}| \ll 
|\tilde{m}_{b_{1}}^{2}+\tilde{m}_{b_{2}}^{2}|$ does not vanish in the infinite 
$\tilde{m}_{t_{1}}$ and $\tilde{m}_{b_{1}}$ limit. However, when the 
three parameters $S_{\tilde{q}}, T_{\tilde{q}}$  and $U_{\tilde{q}}$ are analized
together and the extra condition $|\tilde{m}_{t_{1}}^{2}-\tilde{m}_{b_{1}}^{2}| \ll 
|\tilde{m}_{t_{1}}^{2}+\tilde{m}_{b_{1}}^{2}|$ is included, then the above 
dominant term in $S_{\tilde{q}}$ also vanishes in the infinite squark masses limit as it was expected.

In order to show the decoupling explicitly one must make 
an expansion of $S_{\tilde{q}}, T_{\tilde{q}}$  and $U_{\tilde{q}}$ in powers
 of the proper expansion parameters, which in the (third generation) squarks sector are:
$$ \frac{\tilde{m}_{t_{1}}^{2} - \tilde{m}_{b_{1}}^{2}}
{\tilde{m}_{t_{1}}^{2} + \tilde{m}_{b_{1}}^{2}},
\hspace{0.5cm} \frac{\tilde{m}_{b_{1}}^{2} - \tilde{m}_{b_{2}}^{2}}
{\tilde{m}_{b_{1}}^{2} + \tilde{m}_{b_{2}}^{2}},
\hspace{0.5cm} \frac{\tilde{m}_{t_{i}}^{2} - \tilde{m}_{b_{j}}^{2}}
{\tilde{m}_{t_{i}}^{2} + \tilde{m}_{b_{j}}^{2}},
\hspace{0.8cm} (i,j=1,2).$$
\hspace{0.5cm} In terms of these parameters we get,
\begin{eqnarray}
\displaystyle 
S_{\tilde{q}} &\rightarrow& -\sum_{\tilde{q}} \frac{N_{c}}{18 \pi} \left\{ \left(
\frac{\tilde{m}_{t_{1}}^{2} - \tilde{m}_{b_{1}}^{2}}
{\tilde{m}_{t_{1}}^{2} + \tilde{m}_{b_{1}}^{2}} \right) + 
s_{b}^{2} \left( \frac{\tilde{m}_{b_{1}}^{2} - \tilde{m}_{b_{2}}^{2}}
{\tilde{m}_{b_{1}}^{2} + \tilde{m}_{b_{2}}^{2}} \right) -  
s_{t}^{2} \left( \frac{\tilde{m}_{t_{1}}^{2} - \tilde{m}_{t_{2}}^{2}}
{\tilde{m}_{t_{1}}^{2} + \tilde{m}_{t_{2}}^{2}} \right) \right.\nonumber \\ 
 && \displaystyle \left. -3 c_{t}^{2} s_{t}^{2} \left(
\frac{\tilde{m}_{t_{1}}^{2} - \tilde{m}_{t_{2}}^{2}}
{\tilde{m}_{t_{1}}^{2} + \tilde{m}_{t_{2}}^{2}} \right)^{2} - 
3 c_{b}^{2} s_{b}^{2} \left( \frac{\tilde{m}_{b_{1}}^{2} - \tilde{m}_{b_{2}}^{2}}
{\tilde{m}_{b_{1}}^{2} + \tilde{m}_{b_{2}}^{2}} \right)^{2} \right\}+ 
O\left(\frac{\tilde{m}_{i}^{2} - \tilde{m}_{j}^{2}}{\tilde{m}_{i}^{2} 
+ \tilde{m}_{j}^{2}}\right)^{3} , \\
\nonumber \\
\displaystyle T_{\tilde{q}} &\rightarrow& \sum_{\tilde{q}} \frac{N_{c}}{16 \pi}
\frac{1}{s_{{\scriptscriptstyle W}}^{2} m_{{\scriptscriptstyle W}}^{2}}
\left\{ c_{t}^{2} c_{b}^{2} (\tilde{m}_{t_{1}}^{2} - \tilde{m}_{b_{1}}^{2}) \left(
\frac{\tilde{m}_{t_{1}}^{2} - \tilde{m}_{b_{1}}^{2}}{\tilde{m}_{t_{1}}^{2} + \tilde{m}_{b_{1}}^{2}} \right) + 
c_{t}^{2} s_{b}^{2} (\tilde{m}_{t_{1}}^{2} - \tilde{m}_{b_{2}}^{2}) \left(
\frac{\tilde{m}_{t_{1}}^{2} - \tilde{m}_{b_{2}}^{2}}{\tilde{m}_{t_{1}}^{2} 
+ \tilde{m}_{b_{2}}^{2}} \right) \right. \nonumber \\ 
&& \displaystyle 
-c_{t}^{2} s_{t}^{2} (\tilde{m}_{t_{1}}^{2} - \tilde{m}_{t_{2}}^{2}) \left(
\frac{\tilde{m}_{t_{1}}^{2} - \tilde{m}_{t_{2}}^{2}}{\tilde{m}_{t_{1}}^{2} + \tilde{m}_{t_{2}}^{2}} \right) +  
s_{t}^{2} c_{b}^{2} (\tilde{m}_{t_{2}}^{2} - \tilde{m}_{b_{1}}^{2}) \left(
\frac{\tilde{m}_{t_{2}}^{2} - \tilde{m}_{b_{1}}^{2}}{\tilde{m}_{t_{2}}^{2} 
+ \tilde{m}_{b_{1}}^{2}} \right) \nonumber \\
&& \displaystyle \left. 
+s_{t}^{2} s_{b}^{2} (\tilde{m}_{t_{2}}^{2} - \tilde{m}_{b_{2}}^{2}) \left(
\frac{\tilde{m}_{t_{2}}^{2} - \tilde{m}_{b_{2}}^{2}}{\tilde{m}_{t_{2}}^{2} + \tilde{m}_{b_{2}}^{2}} \right) -
s_{b}^{2} c_{b}^{2} (\tilde{m}_{b_{1}}^{2} - \tilde{m}_{b_{2}}^{2}) \left(
\frac{\tilde{m}_{b_{1}}^{2} - \tilde{m}_{b_{2}}^{2}}{\tilde{m}_{b_{1}}^{2} 
+ \tilde{m}_{b_{2}}^{2}} \right) \right\} \displaystyle + O \left[ \left( \frac{\tilde{m}_{i}^{2} - \tilde{m}_{j}^{2}}
{M_{{\scriptscriptstyle W}}^{2}} \right) 
\left( \frac{\tilde{m}_{i}^{2} - \tilde{m}_{j}^{2}}{\tilde{m}_{i}^{2} 
+ \tilde{m}_{j}^{2}} \right)^{2} \right],\nonumber \\
\\
\displaystyle U_{\tilde{q}} &\rightarrow& \sum_{\tilde{q}} \frac{N_{c}}{12 \pi} 
\left\{ -c_{t}^{2} c_{b}^{2}  \left( \frac{\tilde{m}_{t_{1}}^{2} - 
\tilde{m}_{b_{1}}^{2}}{\tilde{m}_{t_{1}}^{2} + \tilde{m}_{b_{1}}^{2}}
\right)^{2} - c_{t}^{2} s_{b}^{2} \left( \frac{\tilde{m}_{t_{1}}^{2} - 
\tilde{m}_{b_{2}}^{2}}{\tilde{m}_{t_{1}}^{2} + \tilde{m}_{b_{2}}^{2}}
\right)^{2} - s_{t}^{2} s_{b}^{2} \left( \frac{\tilde{m}_{t_{2}}^{2} - \tilde{m}_{b_{2}}^{2}}
{\tilde{m}_{t_{2}}^{2} + \tilde{m}_{b_{2}}^{2}} \right)^{2}\right.\nonumber \\
&&\displaystyle \left. -s_{t}^{2} c_{b}^{2} \left(
\frac{\tilde{m}_{t_{2}}^{2} - \tilde{m}_{b_{1}}^{2}}{\tilde{m}_{t_{2}}^{2} + 
\tilde{m}_{b_{1}}^{2}} \right)^{2} +   c_{t}^{2} s_{t}^{2} \left(
\frac{\tilde{m}_{t_{1}}^{2} - \tilde{m}_{t_{2}}^{2}}{\tilde{m}_{t_{1}}^{2} + 
\tilde{m}_{t_{2}}^{2}} \right)^{2} + s_{b}^{2} c_{b}^{2} \left(
\frac{\tilde{m}_{b_{1}}^{2} - \tilde{m}_{b_{2}}^{2}}{\tilde{m}_{b_{1}}^{2} + 
\tilde{m}_{b_{2}}^{2}} \right)^{2} \right\} + O \left(
\frac{\tilde{m}_{i}^{2} - \tilde{m}_{j}^{2}}{\tilde{m}_{i}^{2} + \tilde{m}_{j}^{2}} 
\right)^{4}. \nonumber \\
\end{eqnarray}
\hspace*{0.5cm}First, we see that in the limit of exact custodial ${SU(2)}_{\scriptscriptstyle V}$ 
symmetry, which corresponds to $\tilde{m}_{t_{1}}=\tilde{m}_{b_{1}} \equiv \tilde{m}_{1}$, $\tilde{m}_{t_{2}}=\tilde{m}_{b_{2}} \equiv 
\tilde{m}_{2}$ and $c_{t}=c_{b} \equiv c$,  $s_{t}=s_{b} \equiv s$, both $T_{\tilde{q}}$ and $U_{\tilde{q}}$
vanish as it is expected, whereas $S_{\tilde{q}}$ goes as,
\begin{equation}
\begin{array}{l}
\displaystyle
S_{\tilde{q}} \rightarrow  \sum_{\tilde{q}} \frac{N_{c}}{3 \pi} c^{2} s^{2} 
\left( \frac{\tilde{m}_{1}^{2} - \tilde{m}_{2}^{2}}{\tilde{m}_{1}^{2} + \tilde{m}_{2}^{2}} \right)^{2} + 
O \left( \frac{\tilde{m}_{1}^{2} - \tilde{m}_{2}^{2}}{\tilde{m}_{1}^{2} + \tilde{m}_{2}^{2}} \right)^{4}
\end{array}
\end{equation}
Second, the above formulae show that the decoupling indeed occurs in the three parameters 
since they go to zero as some power of the proper parameters defined above, which vanish 
in the infinite masses limit. Besides, the decoupling is much faster in $U_{\tilde{q}}$
than in $S_{\tilde{q}}$ and $T_{\tilde{q}}$. These results confirm the numerical analyses performed 
in the literature and agree with the qualitative behaviour discussed in \cite{CHA}-\cite{HA2}-\cite{SOLA}.

However, we would like to emphasize once more that, contrary to most of the 
studies in the literature (with the exception of those on $\Delta \rho$), our results
are model independent and do not make any reference on whether there is or not a 
common effective scale of supersymmetry breaking.

Parallel results for the sleptons sector can be obtained by making the corresponding
replacements in the above formulae \cite{GEISHA}.

\subsection{Neutralinos and Charginos sector:}
\hspace*{0.5cm}Considering together the conditions given for this sector in section 
3.2 we get,
\begin{eqnarray}
\label{eq:Sneuchar}
\displaystyle 
S_{\tilde{\chi}} &=& \frac{-1}{3 \pi} \log \frac{2 \tilde{M}_{2}^{+^{2}}}
{\tilde{M}_{3}^{o^{2}} + \tilde{M}_{4}^{o^{2}}}, \\
\nonumber \\
\label{eq:Tneuchar}
\displaystyle 
T_{\tilde{\chi}} &=& \frac{1}{4 \pi m^{2}_{\scriptscriptstyle W} {\sw}^{2}} 
\left\{ -2{(\tilde{M}_{1}^{+} - \tilde{M}_{2}^{o})}^{2}
\log \frac{\tilde{M}_{1}^{+^{2}} + \tilde{M}_{2}^{o^{2}}}{2{\mu}_{o}^{2}} -
\frac{1}{2} {(\tilde{M}_{2}^{+} - \tilde{M}_{3}^{o})}^{2}
\log \frac{\tilde{M}_{2}^{+^{2}} + \tilde{M}_{3}^{o^{2}}}{2{\mu}_{o}^{2}}\right.\nonumber \\
&-& \displaystyle \left.\frac{1}{2} {(\tilde{M}_{2}^{+} - \tilde{M}_{4}^{o})}^{2}
\log \frac{\tilde{M}_{2}^{+^{2}} + \tilde{M}_{4}^{o^{2}}}{2{\mu}_{o}^{2}} +
\frac{1}{2} {(\tilde{M}_{3}^{o} - \tilde{M}_{4}^{o})}^{2}
\log \frac{\tilde{M}_{3}^{o^{2}} + \tilde{M}_{4}^{o^{2}}}{2{\mu}_{o}^{2}} \right\},\\
\nonumber \\
\label{eq:Uneuchar}
\displaystyle 
U_{\tilde{\chi}} &=& \frac{4}{3{\sw}^{2}} \log \frac{\tilde{M}_{2}^{o^{2}} + \tilde{M}_{1}^{+^{2}}}{2\tilde{M}_{1}^{+^{2}}} +
\frac{1}{3{\sw}^{2}} \log \left[ 
\frac{(\tilde{M}_{3}^{o^{2}} + \tilde{M}_{2}^{+^{2}})(\tilde{M}_{4}^{o^{2}} + \tilde{M}_{2}^{+^{2}})}
{2\tilde{M}_{2}^{+^{2}}(\tilde{M}_{3}^{o^{2}} + \tilde{M}_{4}^{o^{2}})}\right],
\end{eqnarray}
Here the values of the coupling matrices $O_{{\scriptscriptstyle L,R}}, {O'}_{{\scriptscriptstyle L,R}}, 
{O''}_{{\scriptscriptstyle L,R}}$ corresponding to the large neutralinos and 
charginos masses limit have been used.\cite{GEISHA}

With regard to this sector and by looking at 
eqs.(\ref{eq:Sneuchar} - \ref{eq:Uneuchar}) we can conclude that, in the large masses 
limit, the first chargino $\chiu$ and the two first neutralinos
${\tilde{\chi}^{o}}_{{\scriptscriptstyle 1}}$ and ${\tilde{\chi}^{o}}_{{\scriptscriptstyle 2}}$
decouple completely in the $S$ parameter. These are precisely the chargino and
neutralinos, that in the large masses limit become predominantly gauginos. The
decoupling of the other eigenstates $\chid, {\tilde{\chi}^{o}}_{{\scriptscriptstyle 3}}$
and ${\tilde{\chi}^{o}}_{{\scriptscriptstyle 4}}$ in $S$ is not evident at a first sight, 
since it depends on the relative size of the $\chid$ mass and the masses 
of the neutralinos ${\tilde{\chi}^{o}}_{{\scriptscriptstyle 3}}$
and ${\tilde{\chi}^{o}}_{{\scriptscriptstyle 4}}$. However, we have seen in the
above sections, that in the large masses limit, their corresponding squared mass eigenvalues approach
to a common value ${\mu}^{2}$ and, in consequence, the decoupling in $S_{\tilde{\chi}}$ does
finally occur. Notice that this result is not model dependent either, since this common
value ${\mu}^{2}$ is the unique squared mass parameter that is allowed by supersymmetry 
to be present at the  Lagrangian level and do not depend on the particular assumed SUSY 
breaking mechanism. Similarly, in the $T_{\tilde{\chi}}$ and $U_{\tilde{\chi}}$ parameters 
the decoupling occurs exactly if the mass eigenvalues in the large mass limit are considered,
i.e, ${\tilde{M}}_{1}^{+^{2}} \rightarrow M_{2}^{2}, \hspace*{0.1cm}
{\tilde{M}}_{2}^{+^{2}}\rightarrow \mu^{2}, \hspace*{0.1cm}{\tilde{M}}_{1}^{o^{2}}
\rightarrow M_{1}^{2}, \hspace*{0.1cm}{\tilde{M}}_{2}^{o^{2}}\rightarrow M_{2}^{2}$
and ${\tilde{M}}_{3}^{o^{2}}={\tilde{M}}_{4}^{o^{2}} \rightarrow \mu^{2}$.

Finally, we would like to point out that the results for $S, T$ and $U$ of the various sectors are
finite as they must be and the cancellation  of divergences occur between the $\tilde{t}$ and
 $\tilde{b}$ contributions of each generation of squarks, between the $\tilde{\nu}$ and $\tilde{\tau}$
contributions of each generation of sleptons and between the charginos and neutralinos.

\vspace{0.7cm} 
\section{Conclusions}
\label{sec:con}

\hspace*{0.5cm} The computation of the effective action for the standard particles which results
by integrating out all the heavy supersymmetric particles will provide the answer
to the question wether the decoupling of heavy supersymmetric particles in the MSSM occurs
leading to the SM as the remaining low energy effective theory. In this work we have 
shown that the contributions from the heavy sparticles to the two point functions part of 
the effective action can be absorbed into redefinitions of the Standard Model parameters or they 
are suppressed by inverse powers of the heavy sparticles masses.

More specifically, we have proved analytically that the decoupling of squarks, sleptons, charginos 
and neutralinos, at one loop level, in the two-points functions of the electroweak gauge 
bosons takes places. We have considered the limit where
the sparticle masses are all large as compared to the $W^{\pm}$ and $Z$
masses and the external momentum. Notice that we have not assumed exact universality of 
the masses but we have always worked under the plaussible assumption that the differences
 of their squared masses are much smaller than their sums.

Our results for these two-point Green functions in the large SUSY masses limit have 
been presented analiticaly and given in terms of the sparticle masses. Therefore, they are 
general. Namely, they do not depend on the 
particular choice for the soft-breaking terms. In our opinion, it is more convenient for 
the analysis of the phenomenon  of decoupling to use the physical sparticle masses
themselves, being the proper parameters, rather than some other possible mass parameters 
of the MSSM as, for instance, the $\mu$-parameter or the soft-SUSY breaking 
parameters.
     
We have shown that the decoupling of sparticles also occurs in the 
$S, T$ and $U$ parameters, and we have presented explicit formulae for these parameters, 
which illustrate analytically how this decoupling occurs. 

Finally, we have explored to what extent the hypothesis of generation of SUSY masses by 
soft-SUSY breaking terms is relevant for decoupling and we have found instead that the 
requirement of $SU(3)_{\rm c} \times \gs$ gauge invariance of the explicit mass terms by itself is 
sufficient to get it.

\vspace{0.3cm}
{\bf Acknowledgements}\\
\\
M.J.H and S.P wish to acknowledge the organizers of the International Workshop
on Quantum Effects in the MSSM for their kind hospitality during the school. This 
talk is based in a previous work finished in the last October \cite{GEISHA}, which has 
been partially supported by the Spanish Ministerio de Educaci\'on y Ciencia 
under projects CICYT AEN96-1664 and AEN93-0776, and the fellowship AP95 00503301.

\end{document}